\documentclass[10pt,eqsecnum,twocolumn,superscriptaddress]{revtex4}
\usepackage{graphicx}
\usepackage{amssymb}
\def\beq{\begin{equation}}
\def\eeq{\end{equation}}
\def\bea{\begin{eqnarray}}
\def\eea{\end{eqnarray}}
\def\p0{\phi_{0}}

\begin{document}
\title{Insights into Dark Energy: Interplay Between Theory and Observation}
\author{Rachel Bean}  \email[E-mail: ]{rbean@astro.cornell.edu} \affiliation{Dept. of Astronomy, Cornell University, Ithaca, NY 14882}, \author{Sean Carroll} \email[E-mail: ]{carroll@theory.uchicago.edu} \affiliation{Enrico Fermi Institute, Dept. of Physics, and Kavli Institute for Cosmological Physics, Chicago University, Chicago, IL 60637} \author{Mark Trodden} \email[E-mail: ]{trodden@physics.syr.edu} \affiliation{Dept. of Physics, Syracuse University, Syracuse, NY 13244}
\date{\today}

\begin{abstract}
The nature of Dark Energy is still very much a mystery, and the combination of a variety of experimental tests, sensitive to different potential Dark Energy properties, will help elucidate its origins. This white paper briefly surveys the array of theoretical approaches to the Dark Energy problem and their relation to experimental questions.
\end{abstract}
\maketitle

\section{Introduction}
Recent observations suggest that the universe is spatially flat and
undergoing a period of accelerated expansion. In the context of the
standard cosmological picture based on Einstein's general relativity,
this is most readily explained by invoking a new component of smoothly-distributed
and slowly-varying ``Dark Energy'' (DE).

The evidence for this modification is now very compelling indeed, with measurements of a variety of cosmological observables all consistently indicating its presence.  High redshift supernovae observations \cite{Knop03,Riess04} show that previous hypotheses of dust extinction are now inconsistent with the data. Similarly, proposals for photon number density reduction by photon-axion mixing \cite{Csaki02}, although a possible contributor, cannot explain the supernovae, CMB \cite{Spergel03} and large scale structure observations \cite{Tegmark04} completely. The theoretical origin of the observations, therefore, still remains a highly significant mystery, and an area of considerable research activity.

\begin{figure}
\begin{center}
\includegraphics[width=9cm]{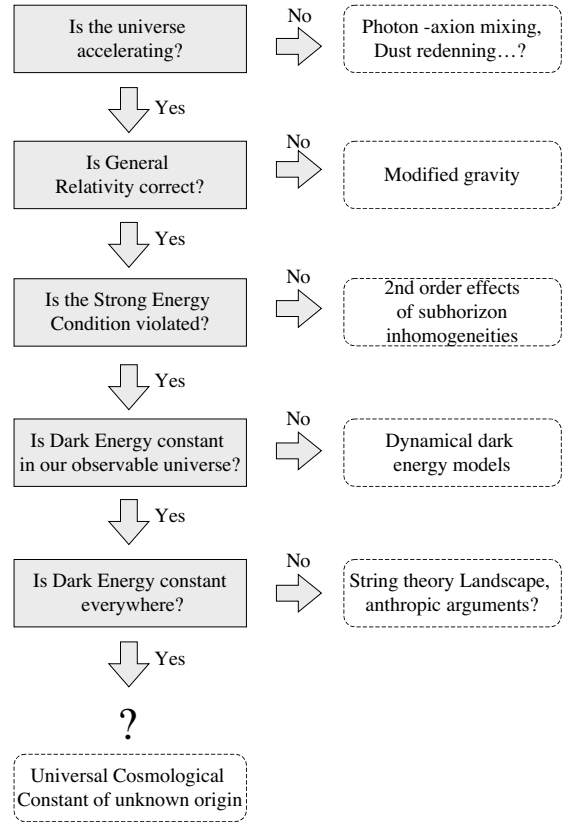} \caption{Summary of the spectrum of current theoretical approaches to the
origin of Dark Energy.} \label{fig1}
\end{center}
\end{figure}

\begin{figure}
\includegraphics[angle=270,width=8cm]{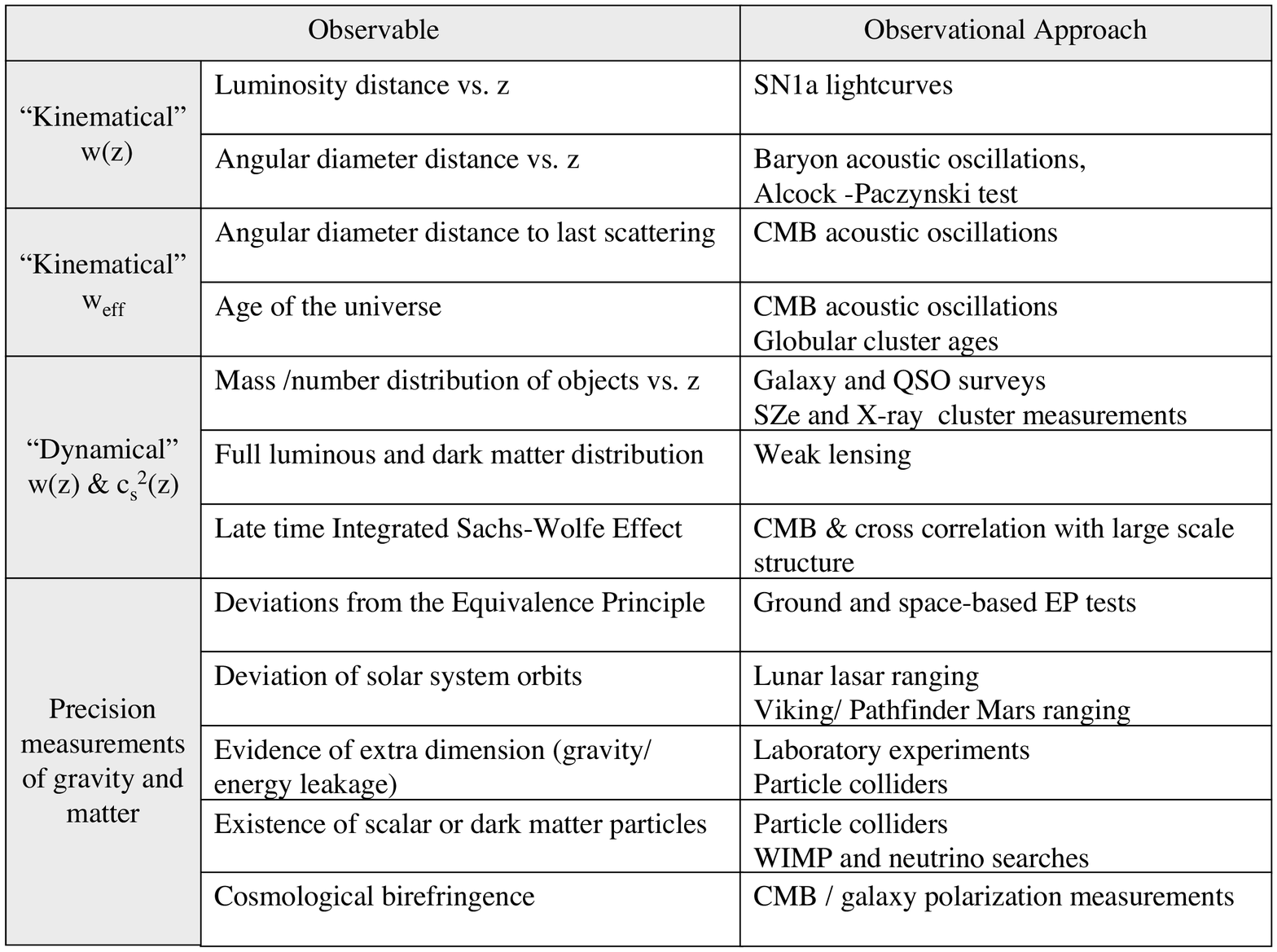}  \caption{Summary of Dark Energy properties tested by observations}
\label{fig2}
\end{figure}

It is the aim of this white paper to highlight the broad spectrum of potential theoretical explanations for the accelerating universe, and the diverse observable signatures, from solar system to horizon scale, relevant to distinguishing between them, summarized in Figs. \ref{fig1} and \ref{fig2}.  We divide the discussion into three basic possibilities:  alternatives to Dark Energy, dynamical Dark Energy, and the cosmological constant. In tandem with observations, a clear view of the theoretical landscape is necessary when assessing future observational strategies. The investigations outlined below therefore play a central role in confronting the Dark Energy problem.

\section{Is there really Dark Energy?}

A homogeneous and isotropic (Friedmann-Robertson-Walker ``FRW") matter-dominated
universe, evolving in accordance with Einstein's equation, will never accelerate.
If we grant the acceleration, then, we must look beyond this framework.
Although Dark Energy is the most obvious and popular possibility, two other
ideas have been investigated:  modifications of gravity on cosmological
scales, and the back-reaction of cosmological inhomogeneities.

General relativity (GR) is very well tested in the solar system, in measurements of the
period of the binary pulsar, and in the early universe, via primordial nucleosynthesis.
None of these tests, however, probes the ultra-large length scales and
low curvatures characteristic of the Hubble radius today.  It is therefore
{\it a priori} conceivable that gravity is modified in the very far
infrared, in such a way that the universe begins to accelerate at late
times.

In practice, however, it is difficult to construct a simple model that
embodies this hope.  A straightforward possibility is to modify the usual
Einstein-Hilbert action by adding terms that are blow up as the scalar
curvature goes to zero \cite{Carroll:2003wy,Carroll:2004de}.  Such theories can lead to
late-time acceleration, but unfortunately typically lead to one of two problems. Either they
are in conflict with tests of GR in the solar system, due to the existence of additional
dynamical degrees of freedom \cite{Chiba:2003ir}, or they contain ghost-like degrees of freedom that seem difficult to reconcile with fundamental theories.  The search is ongoing for versions of this idea that are consistent with experiment.

A more dramatic strategy is to imagine that we live on a brane embedded
in a large extra dimension.  Although such theories can lead to perfectly
conventional gravity on large scales, it is also possible to choose the
dynamics in such a way that new effects show up exclusively in the far
infrared.  An example is the Dvali-Gabadadze-Porrati (DGP) braneworld
model, in which the strength of gravity in the bulk is substantially less
than that on the brane \cite{Dvali:2000hr}.  Such theories can naturally
lead to late-time acceleration \cite{Deffayet:2000uy,Deffayet:2001pu}, but may
have difficulties with strong-coupling issues \cite{Luty:2003vm}.  Most
interestingly, however, DGP gravity and other modifications of GR hold out
the possibility of having interesting and testable predictions that
distinguish them from models of dynamical Dark Energy. Pursuing this
relatively unexplored territory should be a priority of theoretical
research.  A lesson of the investigations carried out to date is that
properties of gravity on relatively short scales, from millimeters to
astronomical units, may be deeply tied to the acceleration of the
universe.

Even in the context of unmodified general relativity, it may be possible
to explain an accelerated universe without Dark Energy by invoking the
effects of inhomogeneities on the expansion rate -- in other words,
perturbations may induce an effective energy-momentum tensor with a
nearly-constant magnitude.  One potentially beneficial aspect of such a
scheme could be to relate the onset of acceleration to the epoch of
structure formation, a coincidence that is otherwise unexplained.

A central question for this approach is whether the feedback of non-linearities into the evolution equations can significantly modify the background, volume-averaged FRW universe and explain the accelerated expansion without the introduction of new matter, or a cosmological constant \cite{Ellis87}.
Key to this issue is that interpreting observations made on a particular scale tacitly also requires the smoothing of theoretical predictions on that scale, and the smoothing operation does not commute with time evolution
 \cite{Buchert00}. The Einstein equations are non-linear, and this non-commutivity means that the FRW equations, for which quantities have been averaged prior to inclusion, will not be the same as the equations obtained by putting in the full inhomogeneous quantities and then averaging the equations.
The difference between the two, over a domain $D$, is a kinematical `backreaction', $Q_{D}$. Along with the, related, averaged spatial Ricci scalar $\langle \mathcal{R}\rangle_{D}$ and the smoothed energy density $\langle\rho\rangle_{D}= M_{D}/V_{D}a^{3}_{D}$, this determines the effective energy density and pressure
 to be used to obtain the evolution equations set by conditions in a domain $D$ \cite{Buchert01} ,
\begin{eqnarray}
\left(\frac{\dot{a}_{D}}{a_{D}}\right)^{2} &=& \frac{8\pi G}{3} \langle \rho\rangle_{D} - \frac{1}{6}\left({\langle \mathcal{R}\rangle_{D}}+ Q_{D}\right)= \frac{8\pi G}{3} \rho_{\rm eff}\\
\left(\frac{\ddot{a}_{D}}{a_{D}}\right)&=&-\frac{4\pi G}{3}\langle \rho\rangle_{D}+Q_{D} = -\frac{4\pi G}{3}\left(\rho_{\rm eff}+3 p_{\rm eff}\right) \label{accel} \ \ \ \ \ \ \ \
\end{eqnarray}
It is claimed that for a large enough backreaction (\ref{accel}) allows the region $D$ to undergo accelerated expansion.

One approach to calculating the  magnitude of the backreaction is to do a perturbative expansion of the metric. A Newtonian analysis does not exhibit acceleration \cite{Siegel05} so the effect must be post-Newtonian, this is puzzling since $G M / (rc^2) \sim v^2/c^2$ is very small for most astrophysical structures. For the post-Newtonian expansion, at first order, the effect on the expansion rate is only $\sim 10^{-5}$, and acts to slow expansion \cite{Rasanen03}.  At second order, for a dust dominated universe, the backreaction behaves entirely like an additional curvature term
 \cite{Russ97,Nambu00}, while in all other environments with different background equations of state (for $-1\le w_{bkg}\le 1/3$) it acts to decelerate the expansion \cite{Nambu02}. Recently, however, Kolb et. al. \cite{Kolb05} have considered sub-horizon higher order corrections to the backreaction, going up to sixth order in a gradient expansion, and suggest that higher order corrections are large enough for the backreaction to generate Dark Energy like behavior, although see \cite{Ishibashi05} for a challenge to this. If the higher order terms are of significance, however, this may imply that a perturbation approach may be inappropriate. Investigations of spherically-symmetric, dust only, Tolman-Bondi-Lemaitre models have concluded that acceleration is possible \cite{Moffat05, Nambu05}. Others, however, studying models within the same class conclude that it is not possible \cite{Alnes05}, or that the analysis suffers instabilities \cite{Iguchi00}.

Future analysis will enable
concrete predictions, establishing the magnitude of the backreaction and a comparison of theoretical predictions against observations. The prospect of
explaining cosmological observations without requiring new energy sources
or a modification of gravity is certainly worthy of investigation.

\section{Is Dark Energy dynamical?}

If Dark Energy exists, it may be phenomenologically described in terms
of an energy density $\rho$ and pressure $p$, related instantaneously
by the equation-of-state parameter $w=p/\rho$.  Covariant conservation of
energy then implies that $\rho$ dilutes as $a^{-3(1+w)}$.  It is
important to stress that $p=w\rho$ is not necessarily the actual equation
of state of the Dark Energy fluid, in the sense that perturbations will
not in general obey $\delta p = w \delta \rho$.  If we have such an
equation of state, however, we may define the sound speed via
$c_{s}^{2} = \partial p/ \partial \rho$.  In order to make sense of
such a picture, we must relate this phenomenology to an underlying
microscopic description.

The simplest candidate for dynamical Dark Energy is an extremely low-mass scalar field, $\phi$, with an effective potential, $V(\phi)$.  If the field is rolling
slowly, its persistent potential energy is responsible for
creating the late epoch of inflation we observe today.

Effective scalar fields are prevalent in supersymmetric field theories and
string/M-theory. For example, string theory predicts that the vacuum expectation value of a scalar field, the dilaton, determines the relationship between the gauge and gravitational couplings. A general, low energy effective action for the massless modes of the dilaton takes the form [assuming a ($-+++$) metric] \cite{Damour94}
\begin{equation}
S = \int d^{4}x \sqrt{-g}\left[ B_{g}(\phi)\frac{R}{2}+B_{\phi}(\phi)K \right] +\sum_{i} B_{i}(\phi)\mathcal{L}_{i}\label{action}
\end{equation}
where $K = \partial_{\mu} \phi\partial^{\mu} \phi $, and $B_{g}$, $B_{\phi}$ and $B_{i}$ are the dilatonic couplings to gravity, the scalar kinetic term and gauge and matter fields respectively, encoding the effects of loop effects and potentially non-perturbative corrections.

In the following sections we describe a range of innovative approaches based on this general action that make distinct  predictions for  a diverse array of observations.

\subsection{Quintessence}

A string-scale cosmological constant or exponential dilaton potential in the string frame translates into an exponential potential in the Einstein frame. Such ``quintessence" potentials  \cite{Wetterich88,Ratra88} can have scaling \cite{Ferreira97}, and tracking \cite{Zlatev99} properties that allow the scalar field energy density to evolve alongside the other matter constituents.
A problematic feature of scaling potentials \cite{Ferreira97} is that
they don't lead to accelerative expansion, since the energy density simply scales with that of matter. It might be that perturbative corrections \cite{Albrecht00} are tuned to generate inflation at the appropriate cosmological epoch.  Alternatively, certain potentials can predict a Dark Energy density which alternately dominates the universe and
decays away; in such models, the acceleration of the universe is just
something that happens from time to time \cite{Dodelson:2001fq}.

Collectively, quintessence potentials predict that the Dark Energy density dynamically evolve in time, in contrast to the cosmological constant. Similar to a cosmological constant, however, the scalar field is expected to have no significant density perturbations within the causal horizon, so that they contribute little to the evolution of the clustering of matter in large-scale structure \cite{Ferreira98}.

The reconstruction of an effective quintessence potential can be determined by measurement of redshift evolution of kinematical observables (those sensitive to the bulk expansion of the Universe): the luminosity distance, measured by supernovae; the angular diameter of the sound horizon at baryon-photon decoupling, measured today by the CMB acoustic peaks, and at earlier epochs via acoustic baryon oscillations in large scale structure correlations; and the linear growth factor inferred from large scale surveys, ratios of weak lensing observables \cite{Jain03, Jain04}, and cross correlation of CMB/ galaxy  \cite{Boughn01,Boughn03,Nolta03,Scranton03, Pogosian05} and weak lensing/ galaxy power spectra \cite{Hu04}.

\subsection{Couplings to ordinary and dark matter}\label{coupled}

A major issue to be confronted by quintessence models is the possibility
of observable couplings to ordinary matter.  Even if we restrict attention
to non-renormalizable couplings suppressed by the Planck scale, tests
from fifth-force experiments and time-dependence of the fine-structure
constant imply that such interactions must be several orders of magnitude
less than expected \cite{Carroll:1998zi}.  One way to evade such constraints
 is to impose an
approximate global symmetry, in which case the quintessence field is
a pseudo-Nambu-Goldstone boson \cite{Frieman:1991tu}; an interesting
prediction of such models is cosmological birefringence, which may be
observed through polarization measurements of distant galaxies and the CMB
\cite{Carroll:1998zi}.  Further improvement
of existing limits on violations of the Einstein Equivalence Principle
in terrestrial experiments would also provide important constraints on
dark-energy models.

Another escape from such constraints can be provided by the ``chameleon''
effect \cite{Khoury04, Brax04}. By coupling to the baryon energy density, the scalar field value can vary across space from Solar System to cosmological scales. Though the small variation of the coupling on Earth allows terrestrial experimental bounds to be met, future gravitational experiments in space such as measurements of variations in the Gravitational Constant with SEE \cite{Alexeev01}, or the  EP-violating two-body experiments on the STEP satellite, will be critical tests for the theory.  A related possibility
involves the coupling of quintessence to neutrinos, with important
consequences for the simultaneous consistency of different neutrino
experiments \cite{Fardon:2003eh}.

Finally, it may be possible that the dynamics of the quintessence field
naturally evolves to a point of minimal coupling to matter.
 In \cite{Damour94} it was shown that  $\phi$ could be attracted towards a value $\phi_{m}^{(x)}$ during the matter dominated era that decoupled the dilaton from matter. For universal coupling, $B_{g}=B_{\phi}=B_{i}$, this would allow agreement with equivalence principle tests and tests of general relativity.  (It is not clear, however, that universal coupling is preferred.)  More recently \cite{Veneziano01}, it was suggested that with a large number of non-self-interacting matter species, the coupling constants are determined by the quantum corrections of the matter species, and $\phi$ would evolve as a run-away dilaton with asymptotic value $\phi_{m}\rightarrow\infty$, and $B_{x}\rightarrow C_{x}+\mathcal{O}(e^{-\phi})$.

In addition to couplings to ordinary matter, the quintessence field may have
nontrivial couplings to dark matter \cite{Anderson:1997un,Farrar:2003uw}.
Non perturbative string-loop effects need not lead to universal couplings, with the possibility that the dilaton decouples more slowly from dark matter than it does from gravity and baryons and fermions. This coupling can provide a mechanism to generate acceleration, with a scaling potential, while also being consistent with Equivalence Principle tests. It can also explain why acceleration is occurring only recently, through being triggered by the non-minimal coupling to the cold dark matter, rather than a feature in the effective potential \cite{Bean01, Gasperini02}. Such couplings can not only generate acceleration, but also modify structure formation through the coupling to CDM density fluctuations \cite{Bean02}, in constrast to minimally coupled quintessence models. Dynamical observables, sensitive to the evolution in matter perturbations as well as the expansion of the universe, such as the matter power spectrum as measured by large scale surveys, and weak lensing convergence spectra, could distinguish non-minimal couplings from theories with minimal effect on clustering.

\subsection{Phantom Dark Energy \& Ghost Condensates}

For the runaway dilaton scenario described in section \ref{coupled}, comparison with the minimally coupled scalar field action
\begin{equation}
S_{\phi} = \int d^{4}x\sqrt{-g} \left[\frac{R}{2}+K-V(\phi)\right]
\end{equation}
shows that the usual solutions require the coupling to gravity, $C_{g}$, and  scalar kinetic term, $C_{\phi}$, to be positive.

Having $C_{\phi}<0$ leads to an action equivalent to a ``ghost'' in quantum field theory, and is referred to as ``phantom energy'' in the cosmological context \cite{Caldwell02}. Such a scalar field model could in theory generate acceleration by the field evolving {\it up} the potential toward the maximum. Phantom fields are plagued by catastrophic UV instabilities, as particle excitations have a negative mass \cite{Carroll03,Cline04}; the fact that their energy is unbounded from below allows vacuum decay through the production of high energy real particles and negative energy ghosts that will be in contradiction with the constraints on UHECR, for example from the EGRET experiment \cite{Sreekumar98}.

Such runaway behavior can potentially be avoided by the introduction of higher-order kinetic terms in the action.  One implementation of this idea
is ``ghost condensation'' \cite{Arkani04}. Here, the scalar field has a negative kinetic energy near $\dot\phi=0$, but the quantum instabilities are stabilized by the addition of higher-order corrections to the scalar field lagrangian of the form $K^{2}$. The ``ghost'' energy is then bounded below, and stable evolution of the dilaton occurs with $w\ge-1$ \cite{Piazza04}.  The gradient $\partial_\mu\phi$ is nonzero in the vacuum, violating Lorentz invariance, and may have interesting consequences in cosmology and in laboratory experiments.

It is worth remembering that observations of $w<-1$ could be misleading, however;  they could, for example, be mimicked by fitting dynamical models with fewer degrees of freedom than required \cite{Maor02}, or by modified-gravity theories \cite{Carroll05}.

\subsection{k-Essence and Unified Dark Matter}

Theories in which the kinetic term in the Lagrangian is not the simple, minimal $K$ are generically called `k-essence' models \cite{Armendariz01}; the effective actions of the ``phantom" and ``ghost condensate" theories fall under this description. Such models can have not only dynamical equations of state, but also clustering properties significantly different from quintessence. Clustering Dark Energy would contribute to density perturbation growth on scales larger than its sound horizon, leading to observable effects in large scale CMB and its correlation with large scale structure and weak lensing surveys\cite{Hu98,Bean04, Weller03,Hu04b}.

Generally k-essence models could have solutions in which shocks are formed, associated with caustics at which the equations of motion are not uniquely defined.   Some k-essence theories avoid this problem, however, such as the Born-Infeld theory,
\begin{equation}
S_{\phi} = \int d^{4}x \sqrt{-g} \left[\frac{R}{2}-V(\phi)\sqrt{1+K}\right].
\end{equation}
 This scalar action has generated interest recently, as a possible ``Unified Dark Matter" theory, in which dark matter and Dark Energy are explained by a single scalar field. The energy density and pressure for the scalar field evolve as $p(\phi)=-V^{2}(\phi)/\rho(\phi)$. At early times, the scalar is effectively pressureless, and in terms of the background evolution, behaves like CDM; at late times the pressure becomes negative and drives accelerative expansion.

The model above is in good agreement with kinematical observations, including the luminosity distance measurements from supernovae and the angular diameter distance from the first CMB acoustic peak \cite{Bento02}. Despite this, the model has been found to be inconsistent with observations when one compares the theories predictions with both kinematic and dynamical data.

In proposing the scalar field as physical, and requiring it to replace CDM and DE one has to also calculate how the scalar field density fluctuations evolve, in order to compare them with density power spectra from large-scale structure surveys. The Born-Infeld theory describes a perfect fluid for which the speed of sound for the scalar field $c_{s}^{2}=1$; as such, the scalar density perturbations resist gravitational collapse, and the scalar alone cannot generate sufficient large scale structure to be consistent with observations.

This is true also for the broader set of phenomenological models including the Born-Infeld action, called ``Chaplygin gases''. Despite being consistent with kinematical observations, they are disfavored in comparison to a $\Lambda$CDM \cite{Sandvik04,Bean03}.

\section{Is Dark Energy the Cosmological Constant?}

Current observational constraints imply that the evolution of Dark Energy is entirely consistent with $w=-1$, characteristic of a cosmological constant ($\Lambda$).  $\Lambda$ was the first, and remains the simplest, theoretical solution to the Dark Energy observations.  The well-known ``cosmological constant problem'' -- why is the vacuum energy so much smaller than we expect from effective-field-theory considerations? -- remains unsolved.

Recently an alternative mechanism to explain $\Lambda$ has arisen out of string theory. It was previously widely perceived that string theory would continue in the path of QED and QCD wherein the theoretical picture contained few parameters and a uniquely defined ground state. However recent developments have yieled a theoretical horizon in distinct opposition to this, with a ``landscape'' of possible vacua generated during the compactification of 11 dimensions down to 3 \cite{Kachru03}.

The ``landscape'' has a SUSY sector with degenerate zero vacuum energy; the non-SUSY sector, however, is postulated to contain $>10^{100}$ or even exponentially greater numbers of discrete vacua with finite vacuum energy densities in the range $\{-M_{p}^{4}, M_{p}^{4}\}$. Arguments based on counting vacua suggest that, in the spectrum of this extensive distribution, the existence of phenomenologically acceptable vacua is plausible. In addition, the non-SUSY vacua are metastable and allow the possibility that a region of space will be able to sample a more extended region of the landscape than if the vacua were stable, through bubble formation.

Given the complexity of the landscape, anthropic arguments have been put forward to determine whether one vacuum is preferred over another. It is possible that further development of the statistics of the vacua distribution \cite{Kobakhidze04,Dine05,Garriga05}, and characterization of any distinctive observational signatures \cite{Hamed05, Barnard05,Tegmark05, Burgess05, Freese05, Freivogel05}, such as predictions for the other fundamental coupling constants, might help to distinguish preferred vacua and extend beyond the current vacua counting approach.

\section{Conclusion}

The ultimate theoretical ambition is to determine the full action and origin for Dark Energy. In light of the `landscape' hypothesis, however, which ultimately may not make a precise prediction for the expected vacuum energy, a more pragmatic question and the one that may offer the most direct link between theory and observation is ``Is the Dark Energy a cosmological constant, or not?'' In this sense we can only conclude DE is a cosmological constant after convincing ourselves that the full plethora of observations have yielded null results.  Absence of time variation in the DE energy density and pressure as measured by supernovae and baryonic acoustic oscillation signatures, no evidence of scale dependent deviations from a $\Lambda$CDM matter power spectra, as might be created by non-minimally coupled quintessence or clustering Dark Energy,  no spatial or temporal variations in the fundamental constants or gravitational-inertial equivalence as predicted by chameleon fields, or perceptible deviations from GR.

This requires a broad array of complementary strategies probing both kinematic and dynamical observables. In the meantime, it is crucially important to
pursue experiments in particle physics, to reveal the microscopic dynamics (SUSY, extra dimensions, or something unexpected) that may bear on the problem of Dark Energy.  A coordinated effort between observers, experimentalists, and theorists will be required to make progress on this most vexing problem in fundamental physics.

\section{Acknowledgements}
The authors would like to thank Eanna Flanagan and Ira Wasserman for useful discussions. The work of MT was supported in part by the National Science Foundation, under grant NSF-PHY0354990, by funds from Syracuse University and by Research Corporation. SC is
supported by the National Science Foundation, the US Dept.
of Energy, and the David and Lucile Packard Foundation.

\end{document}